\documentclass[10pt,journal,compsoc]{IEEEtran}

\ifCLASSOPTIONcompsoc
  \usepackage[nocompress]{cite}
\else
  \usepackage{cite}
\fi 
\usepackage{hyperref}

\ifCLASSINFOpdf
  \usepackage[pdftex]{graphicx}
  \DeclareGraphicsExtensions{.pdf,.png,.jpg,.jpeg} 
\else
  \usepackage[dvips]{graphicx}
  \DeclareGraphicsExtensions{.eps}
\fi 

\ifpdf
  \pdfoutput=1\relax                   
  \usepackage{graphicx}                
  \DeclareGraphicsExtensions{.pdf,.png,.jpg,.jpeg} 
\else
  \usepackage{graphicx}                
  \DeclareGraphicsExtensions{.eps}     
\fi%

\graphicspath{{figures/}{pictures/}{images/}{./}} 

\usepackage{microtype}                 
\PassOptionsToPackage{warn}{textcomp}  
\usepackage{textcomp}                  
\usepackage{mathptmx}                  
\usepackage{times}                     
\usepackage{url}
\usepackage{tabu}                      
\usepackage{booktabs}                  
\usepackage{xcolor}
\usepackage{xspace}
\usepackage{enumitem}
\usepackage{expdlist}
\usepackage{multirow}

\usepackage{flushend}       
\usepackage{graphics}      
\usepackage{tabularx}
\usepackage{ccicons}          

\definecolor{color-measure1}{HTML}{17BECF}
\definecolor{color-measure2}{HTML}{FF9896}
\definecolor{color-measure3}{HTML}{2CA02C}
\definecolor{color-measure4}{HTML}{98DF8A}
\definecolor{color-measure5}{HTML}{FF7F0E}
\definecolor{color-measure6}{HTML}{FFBB78}
\definecolor{color-measure7}{HTML}{9467BD}
\definecolor{color-measure8}{HTML}{C5B0D5}
\definecolor{isoblue}{HTML}{6ED0FF}
\definecolor{isoorange}{HTML}{FCB67A}
\definecolor{isogreen}{HTML}{94D49A}

\definecolor{christoph}{rgb}  {0.9, 0.5, 0.1}

\definecolor{jdf}{rgb}  {0.0,0.8,0.0}

\definecolor{petra}{rgb}  {0.7,0.8,0.0}

\definecolor{tanja}{rgb}  {0.0,0.0,0.8}

\definecolor{revision}{rgb}  {0.25, 0.25, 0.9}

\usepackage[nomessages]{fp}
\newcommand{\maxnum}{3}
\newlength\maxlen
\newcommand{\databar}[2][isoorange]{%
  \settowidth{\maxlen}{\maxnum}%
  \addtolength{\maxlen}{\tabcolsep}%
  \FPeval\result{round(#2/\maxnum:4)}%
  \raisebox{0pt}[0pt][0pt]{\rlap{\color{#1}\hspace*{-1\tabcolsep}\rule[-.25\ht\strutbox]{\result\maxlen}{1.20\ht\strutbox}}}%
  \makebox[\dimexpr\maxlen-\tabcolsep][r]{#2}%
}

\newcommand\system{BitConduite\xspace}
\newcommand{\measure}[1]{\textit{#1}\xspace}

\begin{document}
\title{\system: Visualizing and Analyzing\\Entity Activity on the Bitcoin Network}


\author{Christoph Kinkeldey,
  Jean-Daniel Fekete~\textit{Senior Member, IEEE},
  Tanja Blascheck,
  and Petra Isenberg
\IEEEcompsocitemizethanks{
\IEEEcompsocthanksitem Christoph Kinkeldey is with Inria, France. \protect\\
 E-mail: christoph@kinkeldey.de.
\IEEEcompsocthanksitem J.-D. Fekete is with Inria,  France. 
E-mail: jean-daniel.fekete@inria.fr.
\IEEEcompsocthanksitem Tanja Blascheck is with Inria. France. 
E-mail: tanja.blascheck@inria.fr.
\IEEEcompsocthanksitem Petra Isenberg is with Inria. France. 
E-mail: petra.isenberg@inria.fr.}
\thanks{Manuscript received XXXX XX, 2018; revised XXXXX XX, 2018.}}

\markboth{}{}

\IEEEtitleabstractindextext{
\begin{abstract}
We present \system, a visual analytics tool for explorative analysis of financial activity within the Bitcoin network. 
Bitcoin is the largest cryptocurrency worldwide and a phenomenon that challenges the underpinnings of traditional financial systems---its users can send money pseudo-anonymously while circumventing traditional banking systems. 
Yet, despite the fact that all financial transactions in Bitcoin are available in an openly accessible online ledger---the blockchain---not much is known about how different types of actors in the network (we call them \emph{entities}) actually use Bitcoin. 
BitConduite offers an entity-centered view on transactions, making the data accessible to non-technical experts through a guided workflow for classification of entities according to several activity metrics. Other novelties are the possibility to cluster entities by similarity and exploration of transaction data at different scales, from large groups of entities down to a single entity and the associated transactions. 
Two use cases illustrate the workflow of the system and its analytic power.
We report on feedback regarding the approach and the software tool gathered during a workshop with domain experts, and
we discuss the potential of the approach based on our findings.
\end{abstract}

}






\maketitle

\IEEEdisplaynontitleabstractindextext

\IEEEpeerreviewmaketitle





\IEEEraisesectionheading{\section{Introduction}\label{sec:introduction}}

\IEEEPARstart{B}{itcoin} is a digital pseudo-currency based on strong public cryptography (a \emph{cryptocurrency}) and a payment system~\cite{nakamoto2008bitcoin, Narayanan:2016}. Started in 2009, it challenges several notions of traditional banking and government-regulated currencies and transactions: by using Bitcoin people can bypass traditional centrally governed payment systems. Although it does not have the status of an official currency, it is legal to use virtually everywhere, and a number of countries have officially accepted it as `private money'---with Japan even pushing the use of Bitcoin as a payment system~\cite{Wikipedia:2018}. Millions of users have directly transferred Bitcoin virtual money through its decentralized and permissionless peer-to-peer network while building a large open data source called the Bitcoin \emph{blockchain}: transactions bundled in blocks that form a chain. 
A large amount of ``real'' money has already been invested in infrastructures and global ecosystems around Bitcoin and its market capitalization is estimated as \$180 billion at the time of writing\footnote{\url{http://coinmarketcap.com/currencies/bitcoin/}}. The Bitcoin price skyrocketed in the second half of 2017, leading to large investments by the general public before falling again significantly in 2018. 
Bitcoin, and in particular users' transaction activities, are an important data source to study because little is known about how Bitcoin compares to what is known about fiat currencies. Understanding behavior around the currency might help to explain certain Bitcoin phenomena such as its large volatility. Although other blockchain-based applications have emerged, for example, in health care~\cite{mettler2016blockchain}, insurance business, government services~\cite{Underwood:2016:BBB:3013530.2994581} and other cryptocurrencies such as Ethereum~\cite{wood2014ethereum} or LiteCoin~\cite{Takashima:2018:LUG:3235215} have become popular, Bitcoin is still the dominant application of its kind~\cite{Yli-Huumo:2016}. However, the high amount of transaction data and the common practice of Bitcoin users generating and using many pseudo-anonymous (\emph{pseudonymous}) addresses to send and receive Bitcoin hampers the study of  Bitcoin use. In August 2019 the blockchain received over 300,000 daily transactions, had over 440 million transactions in total, and a much higher number of unique addresses in use. Therefore, a high level of technical expertise is required to extract, store, and analyze Bitcoin transactions that domain experts who are interested in Bitcoin usually do not have. Only few tools exist that lower the threshold of Bitcoin analysis and help with a deeper analysis of Bitcoin data.

\begin{figure*}
  \centering
  \includegraphics[width=\textwidth]{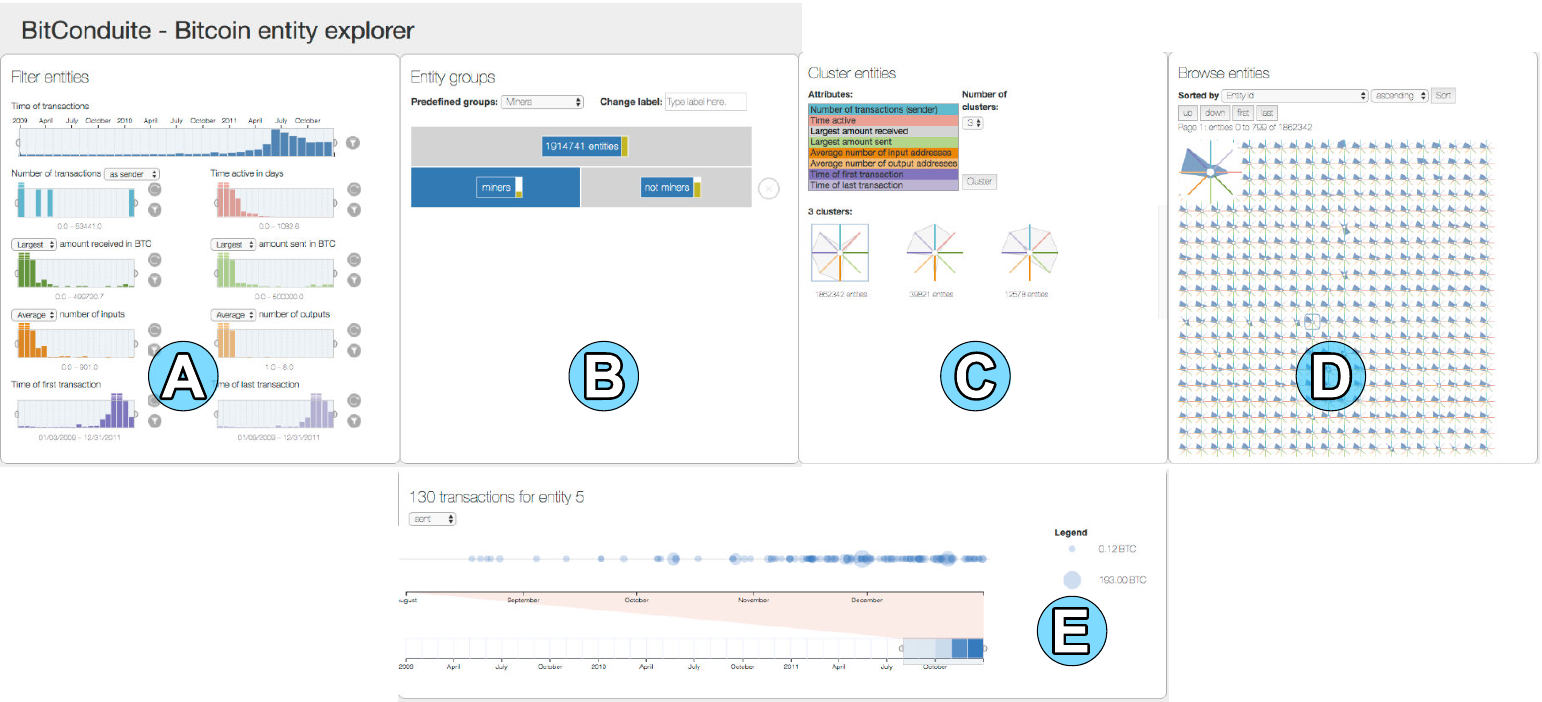}
  \caption{\system's GUI consists of five linked views: (A) filter view, (B) tree view, (C) cluster view, (D) entity browser, and (E) transaction view. Refer to the teaser video for more information (\url{https://vimeo.com/317687395}).}
\label{fig:teaser}
\end{figure*}

We present \system (\autoref{fig:teaser}), a visual analytics tool for the analysis of different types of activities and actor profiles in the Bitcoin network.
It focuses on identifying and characterizing (but not de-anonymizing) individual and groups of \emph{entities}: actors in the network that may be individual users or organizations and services such as exchange platforms.
The contributions of our work include:
\begin{enumerate}[nosep]
\item support for aggregation of raw addresses to \emph{entities} to allow exploratory analysis of meaningful activities over time,
\item a visual exploration tool for human-assisted filtering, classification, and clustering of entities to support data exploration over large-scale time periods,
\item a workflow for systematic classification of entities based on an interactively built decision tree,
\item a discussion on the data back-end we built, the data processing algorithms, and the visual exploration tool we designed to support analysis of the activity of Bitcoin users,
\item and results from a workshop with domain experts that explored potential benefits of our system compared to the status quo.
\end{enumerate}
\system was inspired by regular meetings with experts in economics with an interest in analyzing Bitcoin for their own research. 
This article is a comprehensive extension of a conference poster submission describing an early stage of \system~\cite{kinkeldey:hal-01528605}.

\section{Understanding Bitcoin}
The heart of Bitcoin is a public ledger: the \emph{blockchain}, in which all transactions between users are registered, distributed across the Internet, validated, and maintained. 
There is no overarching structure in Bitcoin that describes users or accounts---transactions are the only description of the system's status~\cite{BitcoinWiki-Transaction:2018}. 
Transaction input and output addresses represent the sender(s) and receiver(s) of a specified amount of Bitcoin. 
The sum of the amounts represented by the input addresses are combined and sent to the outputs. 
As the amounts represented by an address cannot be spent partially, Bitcoin users must, in most cases, combine amounts from multiple addresses to send a certain amount. 
The remainder between the total and the desired amount for the receiver, the change, can be sent back to a sender's address; otherwise, it is offered to the miner validating the transaction.
For example, Bob wants to send 0.015~BTC to Alice. He bought 0.02~BTC using two addresses that hold 0.01~BTC each. To send 0.015~BTC to Alice he uses both addresses with a 0.02~BTC input total. He sends the change of 0.005~BTC back to himself. Like in this example, Bitcoin transactions usually have more than one input address because the amount carried in one address cannot be split up. They also tend to have more than one output address, typically when one output address is the receiver and another one is the change going back to the sender.
 
Due to the basic principle behind Bitcoin, sensitive data about financial transactions are publicly available on the blockchain. 
At the same time the data does not contain personal information about the users. 
Hence, users are pseudonymous but not necessarily anonymous: if an address is linked to information about the user, all transactions involving this address can be traced back in the blockchain, which cannot be deleted or altered. 
For example, if a user publishes a Bitcoin address to receive donations, her or his identity is linked to this address. 
For this reason, it is recommended and common not to reuse addresses but to create a new address for every new transaction.
As many addresses may belong to a single user, address-based analyses of the Bitcoin blockchain are not sufficient if we want to know more about user-based activities.
Actually, an address may belong to an individual, a company, a platform (such as an exchange system), a gambling site, or similar. We call any of them an \emph{entity}, and, in this article, we are interested in \emph{entity-based activities}.



One way to learn about individual entities is to aggregate their addresses. 
A straightforward strategy for this is the \emph{input address heuristic} introduced by Reid \& Harrigan~\cite{Reid:2013}. 
The authors suggested that input addresses of each transaction are likely to belong to the same entity. 
The heuristic is based on the fact that input addresses of a transaction that belong to different entities either have to share private keys or use a special (multi-signature) transaction, both being complex mechanisms unlikely to happen. 
Addresses appearing as input in multiple transactions are linked transitively to the same entity.
This heuristic to aggregate Bitcoin addresses is more conservative than other heuristics (e.g.,Androulaki et al.~\cite{Androulaki:2013}) but one of its advantages is that false positives, i.e., aggregated addresses that in reality do not belong together, are unlikely. 

By using entities instead of addresses it is possible to treat Bitcoin transaction data in a similar way as a bank or credit card account which allows analysts to see all transactions related to the account holder. 
This opens a new perspective on the data and allows a high-level analysis of activity on the Bitcoin network.

\subsection{Linking External Information to Blockchain Data}\label{ssec:external-information}
One way to enrich Bitcoin transaction data is to use lists of identified addresses from external sources to label the entities linked to these addresses.
Known addresses typically belong to exchange services (that trade official currencies into Bitcoin), wallets (online Bitcoin storages), mining pools (that validate transactions and generate Bitcoins), payment services (online merchants that accept to be paid in Bitcoin), or gambling (online gaming). 
Websites such as Walletexplorer~\cite{WalletExplorer:2018} or blockchain.info~\cite{Blockchaininfo-tags:2018} provide lists of known addresses.

In summary, due to the public nature of the Bitcoin system, transaction data are freely available but their expressiveness is limited because of their intrinsic level of anonymity and strategies that users apply to increase this anonymity. Further information on the internals of the Bitcoin system can be found in the literature, e.g., Narayanan et al.~\cite{Narayanan:2016}.

\subsection{Mining New Bitcoins}\label{ssec:mining}
Mining is the process of validating new transactions, building and storing them in blocks, and reaching a consensus on which blocks to add to the blockchain~\cite{BitcoinWiki-Mining:2018}. 
It is a central mechanism of Bitcoin that avoid attacks through manipulation of the blockchain, such as double spending of the same coins. 
Special Bitcoin nodes called \emph{mining nodes} or \emph{miners} have to perform computationally expensive operations to fulfill a \emph{proof-of-work}. 
The node that wins the computing competition mines a new block and earns a reward plus transaction fees voluntarily paid by all senders.
In the beginning of Bitcoin the reward was set to 50~BTC and is automatically halved every 210,000 blocks which happens roughly every four years. 
This mechanism is meant to prevent inflation effects. 
Until today, two halving days took place and the current mining reward per block is 12.5~BTC.
While mining started as a business for individuals, today most of the mining is organized in centralized pools, in which miners collaborate to increase the chance of success and share the revenue~\cite{Bonneau:2015:SOK}.

\section{Related Work}\label{sec:relatedwork}
Despite the growing interest in Bitcoin as a financial and social phenomenon, methods and tools to analyze Bitcoin data only slowly emerge, as well as systematic analyses of the data. 
Some of the reasons could be that the data are not easily accessible via standard APIs~\cite{swan2015blockchain}, that they are hard to interpret because of their pseudonymous nature, and that analysis of the raw data with standard tools (e.g., statistical environments like R) requires non-trivial pre-processing of the data. 
Yet, there are a number of publications reporting on different kinds of analyses of the Bitcoin blockchain. 
We restrict the overview to work on Bitcoin transaction data analysis rather than on technical aspects of the Bitcoin system like its vulnerability. 
In this section, we first introduce approaches related to Bitcoin transaction analysis without visual means. 
After that, we describe approaches using a visual analysis for this task. 
We point out the differences of each approach to ours and motivate the development of \system.

\subsection{Non-Visual Analysis}
Most past analyses of Bitcoin transaction data do not use visualization support (other than displaying the results as charts or node-link-diagrams); they are rather using descriptive statistics and network analysis to describe properties of the Bitcoin blockchain. 
A lot of work has a clear focus on anonymity issues~\cite{Androulaki:2013,Reid:2013,ober2013structure,herrera2015research}, typically by analyzing the structure of the transaction graph and by drawing implications about the anonymity of users. 
Here, we restrict our review on work with the main purpose to analyze user behavior through transaction data.

Ron and Shamir~\cite{ron2013quantitative} conducted an early entity-based analysis of the Bitcoin transaction graph. It yielded results about the characteristics of Bitcoin transactions until mid 2012, for example, the high amount of unused Bitcoins or a number of transactions with large amounts that were all derived from the same transaction in 2010.
Extending this approach, Meiklejohn et al.~\cite{meiklejohn2013fistful} created entities from addresses using two different heuristics and compiled a list of known addresses by crawling mining pools, wallets, and other services. With this information they identified the major players on the network and demonstrated that actual anonymity on the Bitcoin blockchain does not exist. 
As proof-of-concept for their blockchain analysis framework \emph{BitIodine}, Spagnuolo et al.~\cite{spagnuolo2014bitiodine} also used entity-based analysis including known addresses and presented three use cases on illicit activity on the blockchain. 
The insights were novel but analyses with the system require a high level of expertise compared to tools supporting a visual analysis like \system.
As part of their extensive analysis of the first four years of the Bitcoin blockchain, Lischke and Fabian~\cite{Lischke:2016} classified transactions into 13 categories using business tags associated to addresses the owners provided voluntarily. 
They also collected the IP addresses of the transactions.  This way they were able to link them with geo-locations that they determined for a small part of the transactions. 
However, this strategy only led to classification of roughly 50\% of the transactions. 
Maesa et al.~\cite{DiFrancescoMaesa2017} applied graph analysis algorithms to demonstrate the high complexity of the Bitcoin network, to identify major entities, such as the popular ``Mt. Gox'' exchange platform, and to show that there is a concentration of richness in Bitcoin (``rich-get-richer'' effect).
Others followed similar approaches of scraping known addresses from forums and other public posts and linking them to the network of entities~\cite{FlederKP15}. 
Athey et al.~\cite{athey2016bitcoin}, as part of their model for Bitcoin pricing, adoption, and use, suggested a classification tree for user types depending on their activity (e.g., ``one-time user'' or ``long-term frequent transactor''). 
\system extends this approach but facilitates free definition of activity classes instead of fixed classes.

Analyses not using visualization generally reveal structural characteristics of the Bitcoin transaction graph and research anonymity issues. 
However, they answer fixed questions only and analyses are hard to reproduce---issues we address with \system. 
Existing analyses about user activity typically focus on de-anonymizing entities whereas our strategy is to classify entities by their behavior without necessarily de-anonymizing them. 
This way we are not dependent on labels for addresses from public sources that restrict the analysis to entities for which labels exist.

\subsection{Visual Analysis}\label{ssec:related-visual}
Many websites offer simple visual analysis of the Bitcoin blockchain data. 
A popular example is \emph{blockchain.info}~\cite{Blockchaininfo:2018}, a website providing blockchain statistics over time, for instance, the Bitcoin market value, or the number of transactions per block. 
Most of these websites provide information in the form of simple line charts that resemble stock charts and presumably serve a similar purpose: providing information for investors as target users. 
However, there are exceptions that display transactions as circles~\cite{Bitcoin.interaqt:2018}, spheres in a 3D environment~\cite{Bitbonkers:2018}, squares on a 2D map~\cite{Bitnodes:2018}, and bars on a 3D globe~\cite{BitcoinGlobe:2018}. 
Still these kinds of visualizations remain simplistic and restricted to one analytic question, for example, about the largest amounts of Bitcoin transferred in the last minutes.

Only a small number of systems support more complex visual analyses of different Bitcoin characteristics. 
So far, \emph{BitExTract} by Yue et al.~\cite{Yue:2018} is probably the most advanced tool for explorative visual analyses. It supports the analysis of  Bitcoin exchanges, i.\,e., platforms to buy and sell Bitcoin. In four different views, the evolution of transactions between exchanges can be analyzed over time as well as those between exchanges and their clients. Because it focuses on exchanges it serves a different purpose than \system and does not allow an analysis of transactions not connected to an exchange platform.
On a more detailed level, \emph{BitConeView}~\cite{Battista:2015}, displays the traces of specific transactions in a Gantt chart to support the user in detecting suspicious mixing of Bitcoins through the blocks (\emph{taint analysis}). 
Other than \system, it is tailored to one special task and provides insights on a small scale only.
Another visual tool of this kind is \emph{BlockChainVis} by Bistarelli and Santini~\cite{Bistarelli:2017:GBF:3098954.3098972} that shows node-link-diagrams of transactions and allows filtering by block, number of transactions, or the amount of the transaction. 
The basic approach is similar to \system's but it is transaction-centered (not entity-centered), the filtering part is limited in comparison, and advanced processing like clustering is not possible.
McGinn et al.~\cite{mcginn2016visualizing} present a dynamic node-link diagram of transactions between addresses in a visually appealing display. 
The authors identify structures in the graph that may indicate certain types of actors (e.g., commercial platforms). 
However, this approach only displays a short snapshot so no long-term insights are possible. In addition, it shows the raw address-based transaction data and no data processing is possible.

In summary, approaches for visually supported analysis of user activity on the Bitcoin blockchain are rare. In addition, they only cover a small range of functionality. The goal of \system is to offer a generalized, long-term, entity-centered perspective for exploratory analysis of activity that no other approach provides yet.

\section{\system System}
In this section, we describe the components of \system's implementation. 
\system consists of a back end for data preparation and management as well as for high performance data access, and a front end with a graphical user interface (GUI) that consists of five different linked views (\autoref{fig:teaser}). Economist researchers, domain experts who regularly provided feedback, informed the GUI's development. In the following, we describe the components of the system in detail.

\begin{table*}[t]
  \caption{Measures describing entity activity.}\label{tab:measures}
  \begin{tabu} 
  {p{.1\textwidth} p{.05\textwidth} p{.35\textwidth} p{.4\textwidth}}
  \toprule
  \textbf{Measure} & \textbf{Color} & \textbf{Description} & \textbf{Definition} \\
  \toprule
  \measure{num\_txs} & \colorbox{color-measure1}{\phantom{X}} & Number of transactions (as sender / as receiver) & Overall number of transactions per entity. \\
    \midrule
  \measure{time\_first} & \colorbox{color-measure7}{\phantom{X}} & Time of first transaction & Point in time from which an entity was active. \\
  \midrule
  \measure{time\_last} & \colorbox{color-measure8}{\phantom{X}} & Time of last transaction & Point in time until which an entity was active. \\
  \midrule
    \measure{time\_active} & \colorbox{color-measure2}{\phantom{X}} & Time active in days & Duration between the first and the last transaction of an entity. \\
  \midrule
 \measure{amount\_rec} & \colorbox{color-measure3}{\phantom{X}} & Amount received in BTC: smallest / average / largest & Amount of Bitcoin an entity received. \\
  \midrule
  \measure{amount\_sent} & \colorbox{color-measure4}{\phantom{X}} & Amount sent in BTC: smallest / average / largest & Amount of Bitcoin an entity sent. \\
  \midrule
 \measure{num\_inputs} & \colorbox{color-measure5}{\phantom{X}} & Number of inputs: smallest / average / largest & Number of input addresses per transaction. \\
  \midrule
  \measure{num\_outputs} & \colorbox{color-measure6}{\phantom{X}} & Number of outputs: smallest / average / largest & Number of output addresses per transaction. \\
  \bottomrule
  \end{tabu}%
\end{table*}

\subsection{Activity Measures}\label{ssec:activity-measures}
\system is a tool to analyze how Bitcoin is being used, and how its usage has changed over time. 
To systematically describe entity activity we defined eight measures compiled in \autoref{tab:measures}. These are simple descriptive statistical measures that we defined together with the experts who informed our development. For them it was important to work with measures of entity activity that are simple and easy to understand.
The most straightforward one is the number of transactions  (\emph{num\_txs})  in which an entity was involved either as a sender or receiver of Bitcoin. 
This is a measure for the general activity of an entity (``How many times did they use Bitcoin?'').
There are three activity measures related to time: the \hyphenation{time-stamp} timestamp of the first and the last transaction an entity took part in (\measure{time\_first} and \measure{time\_last}), and the time between them (\measure{time\_active}). 
They reflect the temporal aspects of an entity's activity.
The other four measures summarize characteristics of the transactions of an entity, i.e., the amounts received (\measure{amount\_rec}), the amounts sent (\measure{amount\_sent}), the number of inputs of the entity's transactions (\measure{num\_inputs}), and the number of outputs (\measure{num\_outputs}).
Other than the first measures the latter relate to the transactions of the entity and we define the ``smallest'' (minimum), ``average'' (mean), and ``largest'' (maximum) value as entity-related measures (e.g., smallest amount sent).
With this set of measures we are able to describe an entity's activity related to number, time, amounts, and type of transactions. In the future, \system will also be extended to include additional activity measures when other types of activities are analyzed.
In all views of the GUI, the colors shown in \autoref{tab:measures} consistently represent the activity measures.
We use the same color hue for pairs of measures to express that they belong together (e.g., \emph{time\_first} and \emph{time\_last}).

\subsection{Data Acquisition and Preparation}
The first steps when analyzing Bitcoin transactions are data acquisition and preparation. 
The latter is necessary because the raw data only contains low level information on transactions that are difficult to interpret. 
We obtained the raw data by installing the free and open source \emph{Bitcoin Core} client~\cite{Bitcoincore:2017} which downloads the blockchain data (over 200 GB at the time of writing).
The client software provides an API to access the data via a remote procedure call (RPC) interface. 
We used the \emph{python-bitcoinrpc} library~\cite{python-bitcoinrpc:2018} to access the data in Python. 
We then imported the data for blocks and raw transactions into a \emph{MongoDB} database~\cite{mongodb:2018}. 
For further data processing we extracted the transaction data to tables of a column-oriented \emph{MonetDB} database~\cite{monetdb:2018}. 
This step allows fast processing of the raw data that we needed to create the \system's activity measures. 
To accelerate the access further, we wrote the derived data (describing entity activity) into \emph{HDF5} files that are loaded into memory where the server software can access them quickly. 
We use a slicing strategy for fast re-computation of the measures for any time range.
Due to \system's exploratory nature it requires computationally expensive data preparation on the fly, i.e., aggregation of entities and computation of activity measures. 
After some experimentation we decided to opt for an in-memory solution that uses the \emph{pandas} (Python data analysis) library~\cite{pandas:2018} for fast reading and filtering using a \emph{DataFrame} data structure. 

\subsubsection{Entity Creation}
An important characteristic of our approach is that we based the analysis on high-level entities rather than on Bitcoin addresses. 
We aggregated the addresses appearing in the transaction data using the Reid and Harrigan~\cite{Reid:2013} heuristic.
First, we exported address pairs that appear together as inputs of a transaction. 
From them we constructed a graph with the addresses as nodes and their co-occurrence, which are input into a transaction, as links using the \emph{NetworkX} library in Python~\cite{networkx:2018}. 
A UnionFind algorithm yielded all the addresses that are linked to the entities, following the heuristic. The result is a list of addresses for each entity id.

\subsubsection{Known Addresses}
We downloaded and scraped lists of known addresses from public sources such as WalletExplorer~\cite{WalletExplorer:2018}. 
With this information we were able to tag over 70,000 addresses that added context to the analysis with \system. 
Although not the major goal of our approach, our collaborators in economics found some de-anonymization of entities helpful for validating and interpreting results, e.,g., when looking for exchanges or mining pools.

\subsection{Graphical User Interface}
\system's GUI (\autoref{fig:teaser}) consists of five linked views: filter view, tree view, cluster view, entity browser, and transaction view.
They are integrated into a single page web application.
All five views are dynamically updated with every change and can be manipulated independently to facilitate iterative exploration of the data. In the following, we describe the five views and provide more details in the use cases section.

\subsubsection{Filter View}\label{ssec:filter-view}
The filter view (\autoref{fig:teaser}-A) is a dashboard that provides an overview on the temporal distribution of transactions (larger histogram on top) as well as histograms for the activity measures listed in \autoref{tab:measures} (smaller histograms below). 
Initially, the time and value distributions are displayed for all entities in the current data set.
The analyst can filter entities on any of the histograms.
To create a new filter, the desired value range can be selected using a brush on the histogram and/or for more precision, by using the text input fields. 
Pressing the filter button confirms the selection and filters the current set of entities.
For some activity measures
the analyst can switch between smallest, average, and largest value per entity, as explained in \autoref{ssec:activity-measures}.
 
\subsubsection{Tree View}\label{ssec:tree-view}
\begin{figure}
\includegraphics[angle=0, width=\columnwidth]{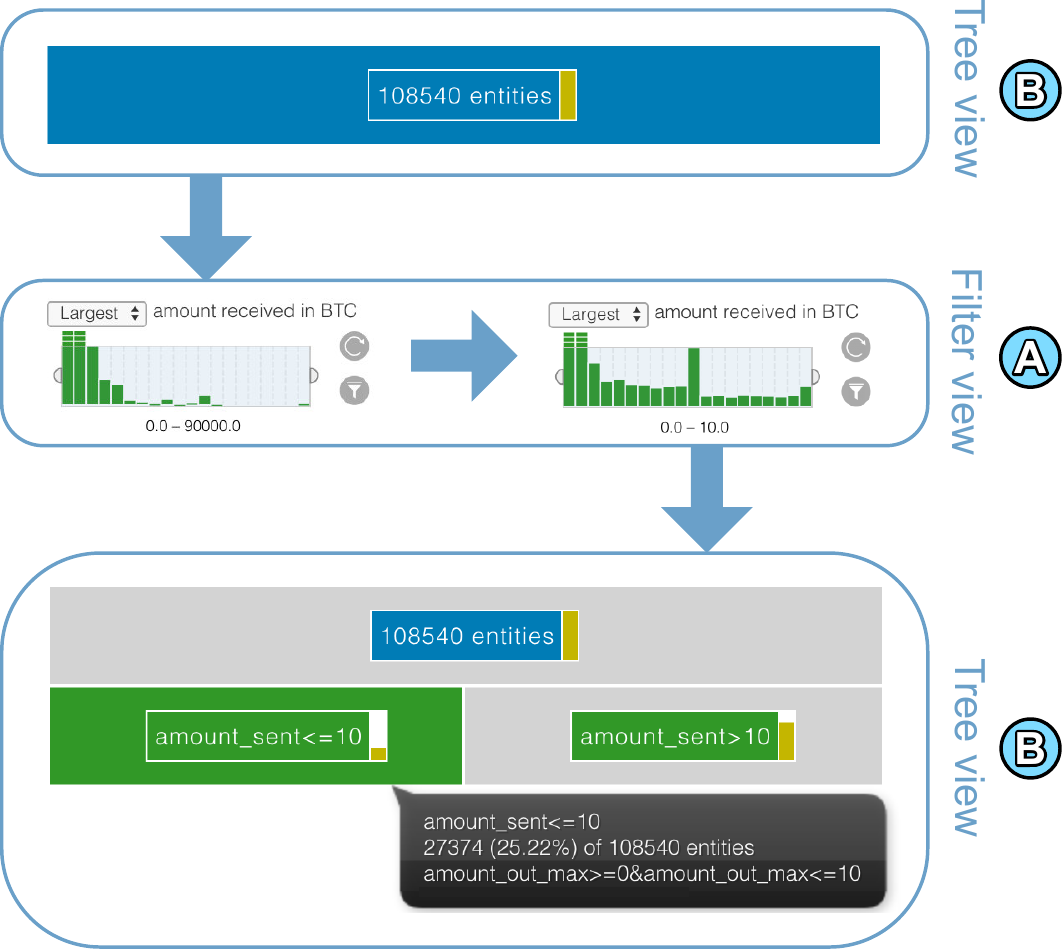}
\caption{Principle of the tree representing classified entities. Filtering the set of 108540 entities yields two subsets shown in the class tree (bottom): the set of entities meeting the condition (left) and the remainder set (right)}
\label{fig:tree}
\end{figure}

The tree view (\autoref{fig:teaser}-B), represents a hierarchy of partitions (we also call them classes and groups) of the entities, visualized as an icicle tree for its compactness.
Initially, only the root node is visible and selected as context on top of the tree, representing the whole set of entities.
Every time a filter is selected in the filter view, it is applied to the selected node. A new row is added in the tree below its parent with two new sets of entities:  those that fulfill the filter condition and the remainder set, see the example in \autoref{fig:tree}: starting with the whole set of entities in the tree view (\autoref{fig:teaser}-B), we see that the histogram for ``largest amount received'' in the filter view (\autoref{fig:teaser}-A) shows a range from 0 to 90,000~BTC. We apply a range filter of 0 to 10~BTC. A new row with two new classes of entities appears in the tree: those that fulfill the filter criterion (left) and the remaining ones (right). The small bars next to the labels signify the relative number of entities per class and tooltips provide detail-in-context. We show the nodes with equal widths and add a glyph to represent the number of entities inside each nodes.   In an initial design, we scaled the size of each node by the number of entities but some groups were small and their nodes became hardly visible.
Clicking on a node selects it as the current context and updates the visualizations. Labeling the nodes and deleting rows is possible as well. That way, the analyst can build up arbitrary entity classification trees and freely switch between the classes to compare their profiles. An export function saves the tree definition to a file to document and exchange the entity class definitions. 

\subsubsection{Cluster View}\label{ssec:cluster-view}
\begin{figure}\centering
\includegraphics[width=0.6\columnwidth]{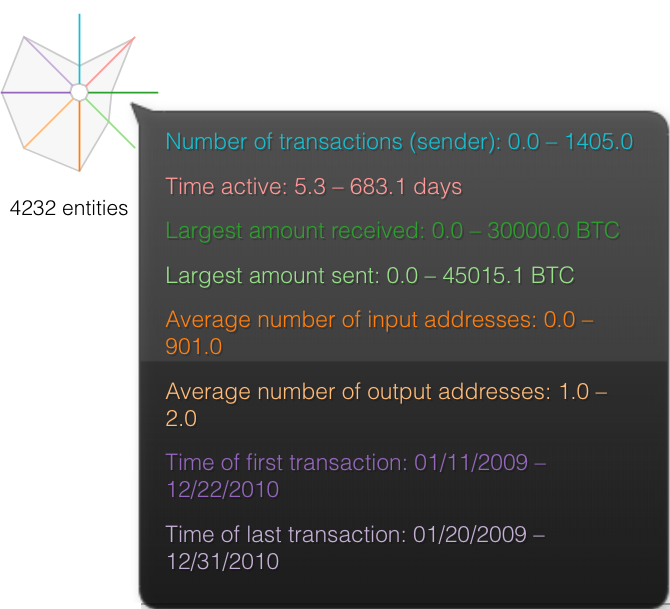}
\caption{Star glyph describing characteristics of a group of entities. Each axis represents the range of values of an activity  measure colored as shown in \autoref{tab:measures}. The shape of the star shows how similar different clusters are. Hovering over the glyph reveals the exact value ranges.}
\label{fig:cluster-glyph}
\end{figure}

The cluster view (\autoref{fig:teaser}-C) provides functionality to automatically create groups of entities sharing similar characteristics. 
Before starting a clustering operation it is necessary to select one or more activity measures and the desired number of clusters.  
For example, an analyst might be interested in entities that sent similar amounts of Bitcoin and were active for a similar amount of time. 
After the clustering is completed, 8-axes star glyphs represent the characteristics of each cluster across all activity measures (\autoref{fig:cluster-glyph}). We chose the star plot over other techniques such as parallel coordinates because star plots allow a representation using small multiples, make the cluster representation relatively scalable without clutter, and allow us to re-use the representation for the display of individual entities.
The glyphs also serve as orientation for the choice of an appropriate number of clusters: if two or more clusters are visually similar and only differ in a measure that is not of interest, the analyst can decide to reduce the number of clusters and restart the clustering.
For our clustering, we use the \emph{k-means} algorithm from the \emph{scikit-learn} machine learning library~\cite{scikit-learn:2018}.

\subsubsection{Entity Browser}\label{ssec:entity-view}
The results of the filtering and clustering steps in \system are groups of entities, and the entity browser (\autoref{fig:teaser}-D) is a browser for those groups. This view helps to get an overview on entities' characteristics and to retrieve information about specific entities. It is not designed to show as many entities as possible but a sample of entities of interest. We chose a $20\times20$ grid of small glyphs showing up to 400 entities at the same time. That way the analyst can compare a large number of entities. Their shapes convey the type of each entity at a glance and reveal similarities and differences between entities. A glyph analogous to the one in the cluster view (\autoref{ssec:cluster-view}) represents every entity. 
Tooltips reveal the values of a glyph for all activity measures. 
When the analyst clicks on one of the glyphs, it is shown magnified and its transactions are displayed in the transaction view. 
The analyst can switch the pages to browse through the entities.  
In addition, the entities can be sorted by different attributes in ascending or descending order, for example, to show the entities with the maximum number of transactions first. 
Or the analyst can determine and compare, e.g., the top ten entities with the maximum amounts received in a transaction.

\subsubsection{Transaction View}\label{ssec:tx-view}
The transaction view (\autoref{fig:teaser}-E) displays a timeline with transactions for a single entity selected in the entity browser. 
This reveals the temporal distribution of transactions and the amounts, answering questions like: did users make transactions within a certain month only or evenly distributed over time? 
The analyst can switch between the transactions in which the entity is involved as a sender or receiver. 
A circle on the timeline represents a transaction with the size of the circle representing the amount sent or received.

\subsection{Workflow}

\begin{figure}
\centering
\includegraphics[width=0.6\columnwidth]{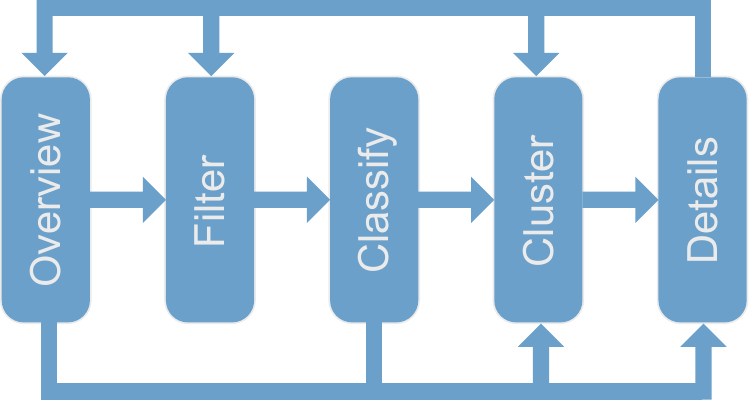}
\caption{Workflow for exploratory Bitcoin activity analysis in \system.}
\label{fig:workflow}
\end{figure}

The workflow for exploratory analysis follows the standard Shneiderman's mantra~\cite{Shneiderman:1996:ETD:832277.834354}: Overview first, zoom \& filter, then details on demand (\autoref{fig:workflow}). In addition, it allows building a hierarchical classification of entities and cluster entities by measures. The GUI supports the following tasks:\\
  \textbf{Overview.} The filter view visualizes the overview of measures for the whole dataset, as well as for the filtered configurations.\\
  \textbf{Filter.} The filter view provides the dynamic queries to filter data and focus on regions of interest over time and over measures.\\
  \textbf{Classify.} The tree view is used to keep track of the filters in a hierarchical way.  Several analyses end up creating multiple \emph{classes} of entities that are given a meaningful name and serve as the outcome of the exploration.\\
  \textbf{Cluster.} The cluster view supports class-level analysis through its interactive clustering method, that helps to find out the value ranges for activity measures for creating meaningful groups. \\ 
  \textbf{Details.} The entity browser allows exploring the entities in detail.\\
The next section showcases some analyses performed using this workflow.

\section{Use Cases}
The main goal of \system\ is a flexible, visualization supported, classification of activities on the Bitcoin blockchain. 
In this section, we present two use cases that demonstrate how an analyst can use the tool to conduct complex analyses in a simple way. 
The first use case is a comparison of activities in the first years of Bitcoin, with a closer look on how activity patterns changed over time. 
The second use case is about the effect of a Bitcoin-specific event, the second ``halving day'' in 2016, on the activities of miners.

\subsection{Classifying Entities: The Early Years}
Bitcoin evolved from a cryptocurrency without any real value to an asset for investment of huge amounts of money that has been running for almost ten years now. 
Because Bitcoin was the first usable system and many similar ones have evolved, it is of interest to analyze how it developed in the first years to predict if or which other cryptocurrencies might be successful.

\begin{figure}
\centering
\includegraphics[angle=0, width=0.75\columnwidth]{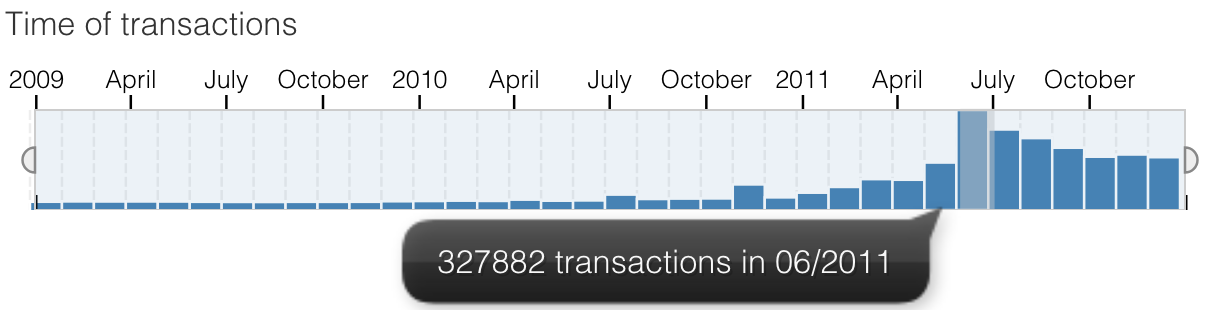}
\caption{Number of transactions per month from 2009 to 2011.}
\label{fig:usecase1-transactions}
\end{figure}

Having a closer look at the years from 2009 to 2011 in \system we can see that around 1.9 million entities have been identified from 2.8 million addresses. 
The transaction view shows that the number of transactions has increased in 2011 with a peak in June representing 328,000 transactions in this month (\autoref{fig:usecase1-transactions}). 
However, how did activity change in the first years?

\subsubsection{One-timers}
\begin{figure}
\centering
\includegraphics[angle=0, width=0.75\columnwidth]{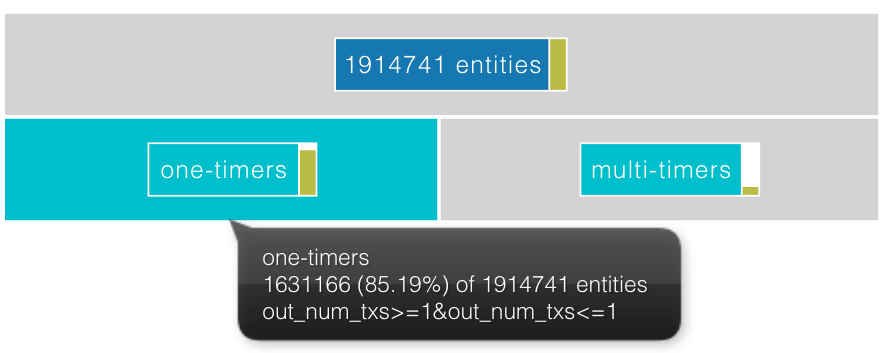}
\caption{The tree view reveals that about 85\% of the entities in the years 2009 to 2011 are one-timers.}
\label{fig:usecase1-tree}
\end{figure}

One research question our domain experts raised was whether Bitcoin is generally used as an investment or actively transferred, for example, to make purchases.
From the filter view we learn that the values for activity measures are highly skewed, especially \measure{amount sent} and \measure{amount received}. 
This is also true for the \measure{number of transactions}: the majority of entities is involved in a low number of transactions. 
To identify all entities with just one transaction (\emph{one-timers}) we define a filter, for which \measure{number of transactions} (as receiver) equals 1. 
The result is a group of entities that only received Bitcoin value once (e.g., by mining or purchasing) and have remained inactive since. 
In the tree view we label this group as ``one-timers'' and the complementary group as ``multi-timers'' (\autoref{fig:usecase1-tree}). 
Until the end of 2011, the majority of all entities (about 85\%) were one-timers. 
There are several possible explanations for this behavior. 
Either people bought Bitcoins and forgot about them, they lost their private key, or they invested and are waiting to sell at a better price. Further filter settings could help with these possible explanations by comparing whether early one-timers transferred their Bitcoins in later years when, for example, the price increased.

Using the time filter we, next, determine that the fraction of one-timers dropped over the years from 95\% (2009) to 87\% (2010) to 85\% (2011). 
This shows that in the first years of existence, most people did not spend their Bitcoins at all. 
The reason is that no exchanges existed before 2010 and although New Liberty Standard defined a first exchange rate\footnote{\url{http://newlibertystandard.wikifoundry.com/page/2009+Exchange+Rate}} in October 2009, Bitcoin's market value was still \$0 in practice. 
The number of people was limited and there was no real commercial opportunity to spend Bitcoins. 
The first official payment in the real world was the purchase of a pizza on May 22, 2010, called ``Bitcoin Pizza Day'' thereafter\footnote{\url{https://en.bitcoin.it/wiki/Laszlo\_Hanyecz}}.

\subsubsection{Most Active Entities}
\begin{figure}\centering
\includegraphics[angle=0, width=0.7\columnwidth]{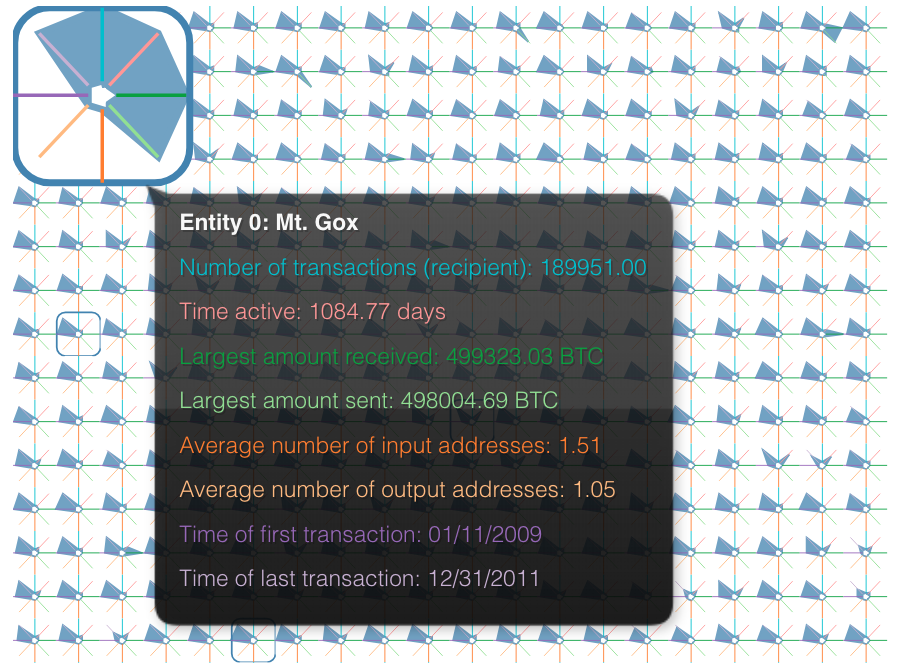}
\caption{The most active entity is Mt. Gox, an early exchange platform.}
\label{fig:usecase1-mostactive_result}
\end{figure}

To identify entities with the highest activity in the early years, we switch to the ``multi-timers'' class by clicking on the associated node in the tree view. 
Looking at the filter view we notice that the \measure{number of transactions} per entity ranges from 0 to 57,384 (as sender) and 1 to 189,951 (as recipient). 
We use the cluster view to cluster the multi-timers by four activity measures: \measure{number of transactions}, \measure{time active}, as well as largest \measure{amount received} and \measure{amount sent}. 
We obtain three clusters: one active cluster with over 35,000 entities and two clusters of entities with less activity with differences in \measure{number of transactions} and average \measure{number of inputs}. 
Because we can tell from the glyphs that the two first clusters are similar we reduce the number of clusters to two and restart clustering. 
The result is now a small cluster of active entities (39,043) and a large one with entities of low activity (244,532). 
By clicking on the glyph representing the first cluster we load the 39,043 active entities into the entity browser. 
Sorting by \measure{number of transactions} yields four entities that are more active in relation to the others.
The most active entity is tagged as ``Mt. Gox'', the Bitcoin exchange platform that started in June 2010\footnote{\url{https://en.wikipedia.org/wiki/Mt.\_Gox}}~(\autoref{fig:usecase1-mostactive_result}).

This simple analysis shows that from their first appearance, big platforms have dominated the blockchain. This is a phenomenon that is still valid nowadays~\cite{harrigan2016unreasonable}.


\subsection{Analyzing A Significant Event: Halving Day 2016}
This second use case deals with the phenomenon of Bitcoin mining. 
The mining market has changed over the years and has become more centralized. Nowadays, large mining pools dominate the market~\cite{Bonneau:2015:SOK}. 
This is of interest for researchers in economics who examine the mining market and its impacts on the Bitcoin system. 
\emph{Halving days} (when mining rewards are cut by 50\%
) are deemed important because they tend to have great impact on the Bitcoin price and the mining behavior.

\begin{figure}\centering
	\includegraphics[angle=0, width=0.75\columnwidth]{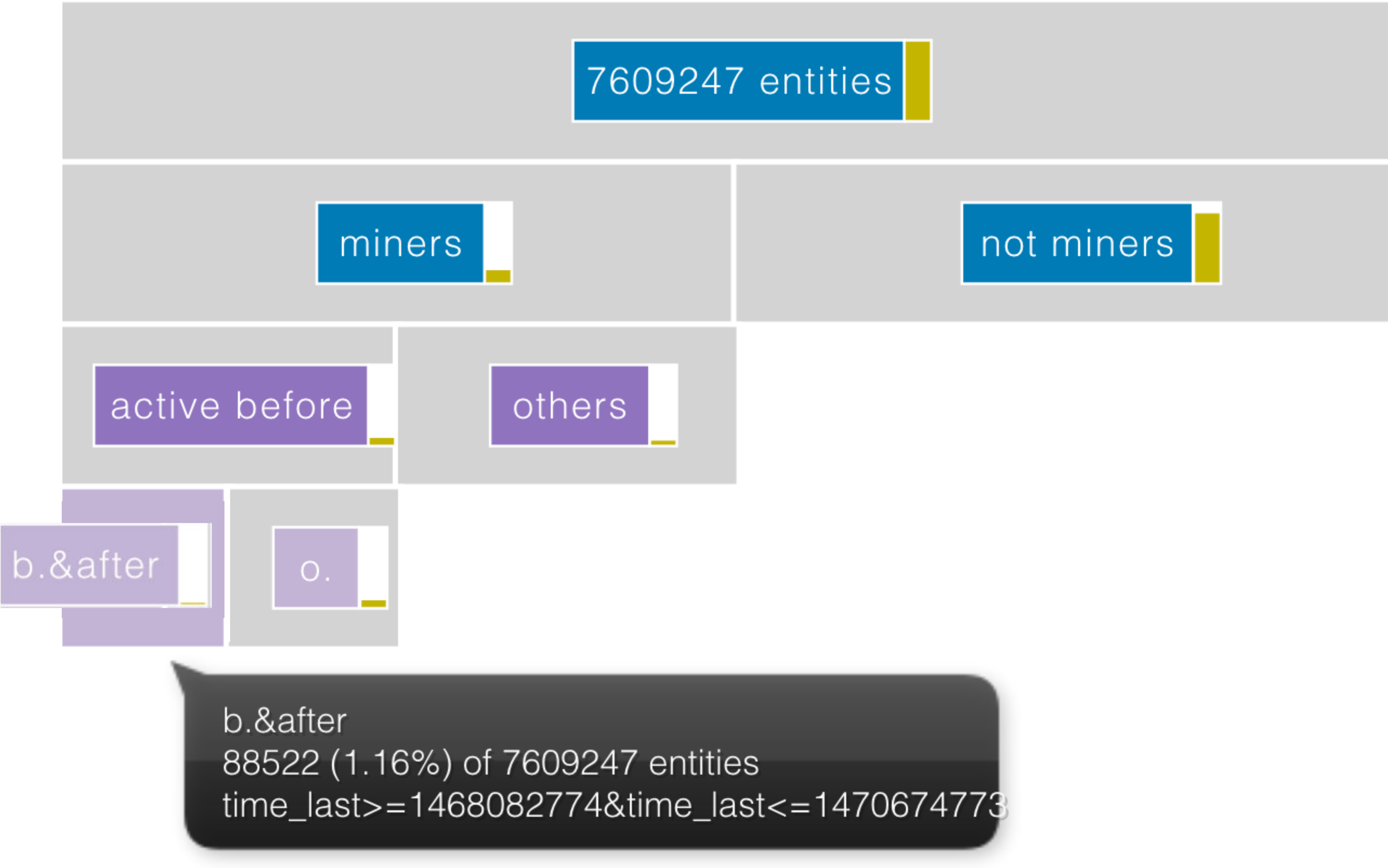}
	\caption{The group of entities (``b.\&after'') that were active as miners before and after the halving day.}\label{fig:usecase2-tree}
\end{figure}

To see effects of the last halving day on July 9, 2016, we take a closer look at mining behavior during the months before and after this day. 
It is known that the Bitcoin price remained relatively stable after this day\footnote{\url{https://charts.bitcoin.com/btc/chart/price}} but not much is known about the change in mining activity. A question our economics experts found particularly interesting was if the same entities successfully mine Bitcoin after the halving day or if some give up due to decrease in profit.

To answer this question we focus on transactions from June 9, 2016 to August 8, 2016, that is, 30 days before and 30 days after the halving day. 
Looking at the time chooser and the tree view, we learn that during this time period, $7.6$ million entities took part in about $13.8$ million transactions (as sender, recipient or both). 
The histograms in the filter view reveal that although the \measure{number of transactions} ranges from 0 to 1,075,326 (as sender) and 1 to 3,074,401 (as receiver) the vast majority of entities took part in a small number of transactions.

\subsubsection{Identify Miners}
Miners can be identified as receivers of the ``coinbase'' transaction, which is the first transaction in each block that rewards them for mining the block~\cite{Narayanan:2016}. 
It does not have a regular input address like other transactions. 
That is why we can identify entities that were involved in mining activities by applying a ``zero input'' filter (smallest recorded \measure{number of inputs} = 0). 
In the classification tree we label them as ``miners'' and the complementary group as ``not miners'' (\autoref{fig:usecase2-tree} second row). We learn that 1,050,441 (roughly 13.8\% of all) entities in this time belong to the ``miners'' group. 
Using the time chooser we find out that the \measure{number of transactions} 
after the halving day is about 16\% lower than in the month before. 
Does this indicate less mining activity after the halving day?

\subsubsection{Activity of Miners around the Halving Day}
To get a better overview about the activities of the miners we cluster them by number of transactions as recipient and the time they were active in the selected time period. Starting with three clusters, the cluster view shows a large group of entities with low activity (905,732 / 86.2\%), a smaller group of high activity (115,373 / 11.0\%) and a tiny one in between (29,336 / 0.03\%). We decide to cluster again with only two groups because the two clusters with low activity are similar and mainly differ in the average number of input addresses, which is not of interest here. 
The result is a large cluster (922,245 / 87,8\%) with low activity and a much smaller one with high activity (128,196 / 12,2\%). 
Looking at the glyphs in the cluster view and retrieving the exact values by hovering over them reveals that, beside the number of transactions, the two groups mostly differ in terms of the range of amounts received and sent and the number of input addresses per transaction.

One hypothesis regarding the halving day is that some miners stop their activity afterwards because of the much lower reward. 
To examine this we need to find miners that were active in the weeks before the halving day and remained active afterwards. 
As a first step, we apply a filter by \measure{time of first transaction} before the halving day yielding a group of 7.9\% of all entities we name ``active before''. We further filter this subgroup setting the \measure{time of last transaction} filter to the time period after the halving day. 
This yields the group of miners that were active before the halving day and remained active afterwards (1.2\% of all entities). 
As a complementary group we get those that ``gave up'', i.e., they were active before but not in the 30 days afterwards (6.7\% of all entities)~(\autoref{fig:usecase2-tree}).

This analysis shows that a large fraction (roughly 85\%) of the miners that were active before the halving day did not receive mining rewards in the following 30 days. 
The reason could either be that they stopped mining after the halving day because the lower reward made their work no longer profitable, or that they still took part in the mining competition but were simply not successful.
The blockchain does not provide information about this and without further information we can only speculate about the reasons. 
However, indeed, the number of miners seems to have decreased significantly after the halving day.

Looking at the entity browser lets us identify the most active entities before the halving day that remained active after. 
The transaction view shows that one of the most active miners from before the mining day was inactive during the 30 days after the halving day. From this observation we can conclude that the halving day had a relevant impact on mining behavior. 

The use cases showed that \system can support a range of complex explorations and insights. Still, we have to keep in mind the uncertainty inherent to our method due to the heuristics used for extracting entities from addresses.
At the same time, the lack of ground truth prevents us from estimating this uncertainty.

\section{Expert Workshop}

We conducted a workshop with Bitcoin experts to receive feedback on how \system supports exploratory analysis of entity behavior on the blockchain based on real analysis questions. We first describe the setup of the workshop and then report on the results.

\subsection{Setup, Procedure, and Data Collection} 
We set up six workstations (standard desktop computers with Windows 10 / Ubuntu and Chrome / Chromium web browsers) in a large shared office in our lab. 
Participants interacted with \system running in the web browser while data was delivered from a database and web server in our local network. We provided participants with data about the early years of Bitcoin from 2009--2011. Three co-authors of this paper were present at the workshop to take notes, pictures, and help with participants' questions. 

Before the start of the workshop, participants answered a questionnaire on demographics and their experience with Bitcoin. They then proceeded to a 30 minute training phase in which we presented the system using a slideshow and gave participants three hands-on exercises to complete. We answered participants' questions throughout the training. We stopped the training after participants confirmed that they understood the \system workflow.
 
After the training, participants filled out a second questionnaire, in which we asked them to write down questions they wanted to explore during the free exploration phase. Next, participants began the 60 minute free exploration phase during which they used \system to answer their own questions about the Bitcoin blockchain. Throughout the free exploration phase the three experimenters answered questions about \system if help was needed. The last questionnaire, after the free exploration phase, contained questions about \system to receive more structured feedback about its usability. 
All three questionnaires are part of the supplemental material, as well as the introduction slides to \system that contain the hands-on-exercises.

During the free exploration, we recorded participants' actions with the tool, i.\,e., all events such as filtering, clustering, defining and selecting entity groups in the tree view, sorting, and selecting entities in the entity browser, as well as switching the mode (input / output) in the transaction view. {Similar to other researchers~\cite{Blascheck2016c,Brown2014,
Guo2016}, 
we used the interaction logs to learn about behavior and reasoning processes trying to understand how our experts arrive at insights during the study. Collecting  interaction data allowed us to analyze which views participants interacted with the most and if they used \system the way we intended (\autoref{fig:workflow}).

\subsection{Background Information}
We invited six participants, most of them with an economics background, to a half-day workshop in our lab. Five participants were male, one was female, and their age ranged from 25 to 51 years (average: 34.7 years). 
Among them, two participants were students (PhD in Economics and MSc in Acturial Finance) and the other participants were professional researchers: a senior researcher, research engineer, assistant professor, and associate lecturer.

We also asked participants about their familiarity with Bitcoin, visual data representations, clustering, and statistics; the results are summarized in \autoref{tab:familiarity}. 
All participants confirmed to have been working with Bitcoin data, specifying their experience between one month and five years (average: 1.9 years, one participant did not provide this information).  We also asked what kind of Bitcoin data the experts worked with, and research questions they studied. Overall, participants were interested in many different topics regarding Bitcoin:
\vspace{-.5ex}\begin{description}[\compact\setlabelphantom{P6}]
\item[P1] \textit{understand what kind of people use Bitcoin and why}
\item[P2] \textit{model the market for mining and proof-of-stake}
\item[P3] \textit{find out if exchanges have liquidity issues and hat kind of smart contracts people use}
\item[P4] \textit{define a sentiment index on Bitcoin using alternatives data}
\item[P5] \textit{answer cryptography questions about robustness}
\item[P6] \textit{study the dynamics of miners' pools}
\end{description}



\begin{table}[tb]
  \centering
  \tabulinesep=2pt
	\vspace{3mm} 
	\caption{Answers of participants to background questions (on 1--5 Likert scales).}
        \scriptsize
  \begin{tabu}{@{}p{.63\columnwidth}lrl@{}}
	\toprule
	Question & 1 & avg & 5\\
	\midrule
	Rate your understanding of how Bitcoin works internally&no&\databar[isoblue]{3.8}&expert\\
	How carefully do you follow news related to Bitcoin? & never &\databar[isoblue]{3.8}& daily\\
	I use visual representations of data during my work & never &\databar[isoblue]{2.3}& daily\\
	I create visual data representations myself & never & \databar[isoblue]{2.5}&daily\\
	I feel confident interpreting basic charts&disagree & \databar[isoblue]{3.7}&agree\\
	I feel confident interpreting k-means clustering results&disagree& \databar[isoblue]{3.3}&agree\\
	I feel confident interpreting basic summary statistics&disagree& \databar[isoblue]{4.5}&agree\\
 \bottomrule
  \end{tabu}
\label{tab:familiarity}
\end{table}


\subsection{Results}

In this section we report on the results from the research questions, interaction logs, and user experience questionnaire.

\subsubsection{Research Questions}


Before the free exploration, we asked participants to write down questions they would like to explore with \system.
Participants wrote down 18 questions. We coded and clustered these questions into three general areas (linking, entities, trends) and five more specific topics (linking, single person entities, specific entities, behavior trends, temporal trends). \autoref{tab:questions} lists questions and categories. 


\begin{table}[tb]
  \centering
  \footnotesize
  \tabulinesep=2pt
	\caption{Example questions participants wanted to explore with \system during the free exploration phase. Questions that can be answered with \system are marked with an asterisks (*). The full set of questions can be found in the supplemental material.}
  \begin{tabu}{%
    @{}X[1.5,l,p]X[3,l,p]@{}
	}
	\toprule
	\textbf{Category} & \textbf{Questions}\\
	\midrule
		\multirow{1}{.4\columnwidth}{Exploring specific entities (known entities)} 
		& Does Kraken have liquidity issues? (P2)\\
		& How many [B]itcoins does Satoshi own? (P2)\\
	\midrule
		\multirow{1}{.4\columnwidth}{Exploring specific entities (other entities)}
		& *Do [entities] [exchange] money with the same people? (P1)\\
	 	& Who are the 10 main owners of Bitcoin and how much [do] they own? (P3)\\
	\midrule
	\multirow{1}{.4\columnwidth}{Linking and rela\-tion\-ships between entities}
		& *Do [entities] send small amounts to 1 person (or the opposite = large amount[s] to multiple)? (P1)\\
		& Do mining pools interact with each other directly? (P1)\\
		& *Explore and link multiple entities to a single one based on its behavior~(P1)\\
	\midrule
	 \multirow{1}{.4\columnwidth}{Exploring trends (behavior)}
	 	& *Which factors affect pools of miners dynamics? (P1)\\
	 	& *What happened to BTC exchange platforms during trouble periods? (P6)\\
	 \midrule
	 \multirow{1}{.4\columnwidth}{Exploring trends (temporal)}
	 	& *Are there daily users that are non-miners / non-professionals? (P1)\\
	 	& Is there any seasonality in the use of BTC? (P5)\\
 \bottomrule
  \end{tabu}
\label{tab:questions}
\end{table}

\subsubsection{Tool Use and User Experience}
Next, we discuss results on how participants used \system and their experience of trying to answer their own research question(s).

\begin{figure}
\centering
\includegraphics[width=\columnwidth]{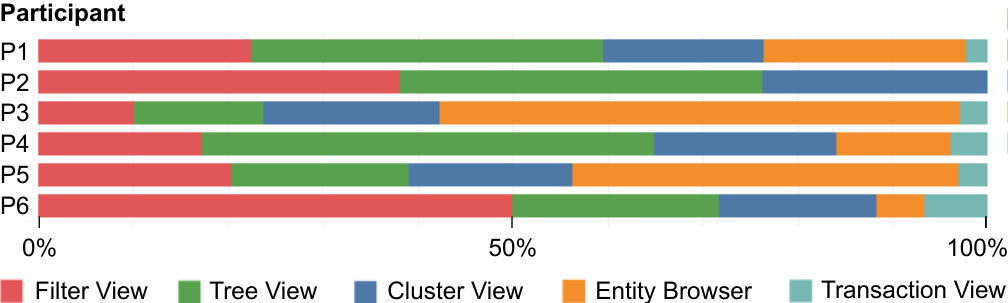}
\caption{Relative proportion of mouse-click interactions each participant executed on the main views of \system.}
\label{fig:logs-interaction}
\end{figure}

\noindent\textbf{Interaction Logs.}
Interaction logs for each participant provide a first approximation of which system features they did or did not use. The logs consisted of a list of events per participant: the type of event (``interaction''), the timestamp, the object that was interacted with (``timechooser\_filter''), and the value (e.,g., filter criteria). We extracted the number of times participants executed an event (through a mouse click), compared to the overall number of clicks, and compiled these proportions in an overview (\autoref{fig:logs-interaction}).

Given that participants were interested in a large variety of different research questions it is not surprising that \autoref{fig:logs-interaction} shows a large variety in the proportions of interactions for each participant. While 5 out of 6 participants interacted with all views, participant~P2 did not interact with the entity browser and the transaction view at all. The reason might be that this participant had questions (see \autoref{tab:questions}) for which detailed data on single entities and transactions was not useful. The filter view exhibited a large variety in the proportion of interactions (10\%--50\%); similarly for the entity browser (0\%--55\%). Overall, the transaction view was interacted with the least. There are several possible explanations for its diminished use: research questions may not require studying individual entities and their transactions; the view was at the bottom of the page; and we did not log all interactions with the timeline (e.g., changing the time range).

The interaction patterns show that all participants exhibited repeated sequences of switching between the filter and tree view (filter $\rightarrow$ tree $\rightarrow$ filter $\rightarrow$ tree $\rightarrow$ filter).
This is not surprising because the filter view modifies the tree view but does show that our choice to place them visually next to each other worked well to reinforce the connection. Looking at our intended workflow for \system (filter $\rightarrow$ tree $\rightarrow$ cluster; \autoref{fig:workflow}), we find that all participants used this sequence except P4. This participant exhibited technical difficulties and long delays and, therefore, his/her workflow was interrupted. Another typical sequence participants P3, P4, and P6 applied was filter $\rightarrow$ cluster $\rightarrow$ entity browser, meaning that they skipped modifying the tree view and just used the automatically selected group. From this data, we conclude that the participants used \system the way we intended and participants used similar interaction strategies to analyze the data.

To complement the interaction log data, we also asked participants to write down their exploration strategies. Two participants (P4, P5) mentioned that they followed the tutorial instructions.
Participant~P1 used the filter view to find specific entities and groups as well as explored clusters and used these to drill down into the transactions of one entity.
Participant~P6 tried to focus on specific entities they had already identified (e.g., Bitminter and Mt. Gox). These strategies are reflected in the interaction logs. 
Participant~P2 and P3 did not answer this question.



\noindent\textbf{User Experience.}
In our exit questionnaire, we asked participants to rate \system regarding different aspects (\autoref{fig:results-questionnaire_tool}). 
No participant found the system unnecessarily complex. One participant agreed and one was neutral that they would need further support to use the tool. 
Similarly, one participant agreed with the statement that they needed to learn many things before its use. All but one participant were confident with using the system, found it easy to use, its functionality well integrated, and would use \system more frequently. We also asked participants what they saw as the strength of the tool and whether they had suggestions for improving \system.
Participants answered that one strength of \system was the \emph{``exploration of huge amounts of data''}~(P1); that it is \emph{``easy to understand and to use''}~(P3); it shows \emph{``a lot of data and information on screen''}~(P4), that it has \emph{``a workflow which makes sense''}~(P4); it has \emph{``[m]any small charts that help having an overall visualization''}~(P5); and \emph{``[i]t provides cluster analysis on already identified entities''}~(P6). 

Participants also mentioned that \system could be improved by adding \emph{``\ldots filters based on frequency\ldots''}~(P1); \emph{``a geographical origin for all transaction to get a map''}~(P3); \emph{``when presenting the tools, tell a discovery story''} and \emph{``explain k-mean''}~(P4); as well as adding a \emph{``network visualization''} and \emph{``the possibility to export data''}~(P6). These results are encouraging for future development.

\begin{figure}[tb]
	\includegraphics[angle=0, width=\columnwidth]{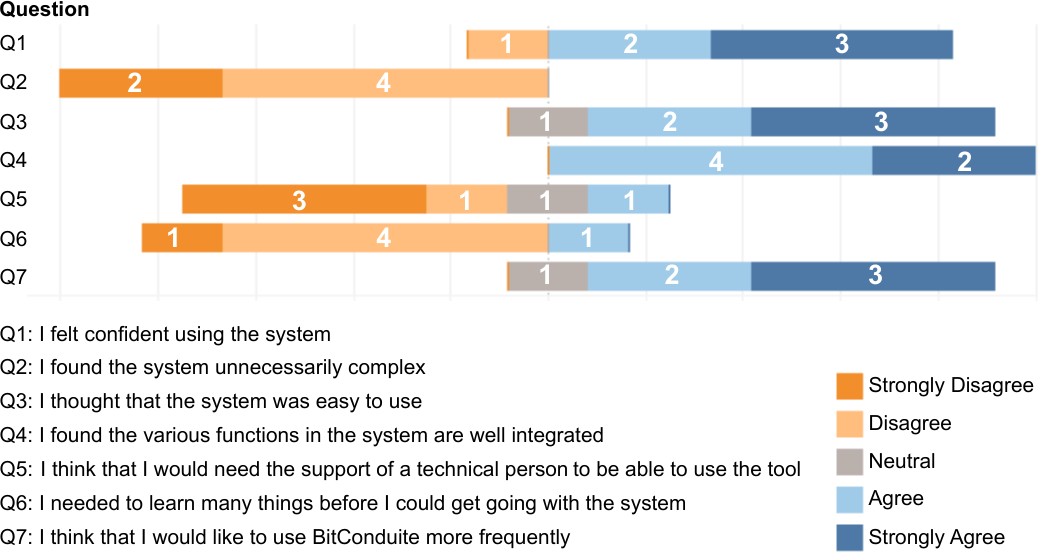}
	\caption{Participant response counts to the questions concerning the usability of \system.}\label{fig:results-questionnaire_tool}
\end{figure}

\noindent\textbf{Answering Research Questions.}
We received eleven answers in the exit questionnaire on whether \system helped participants to answer their questions. If participants had multiple research questions, then some participants mentioned which of the research questions they could answer or not. 
Five answers showed that participants were able to answer their questions. However, six responses were negative as participants were not able to find an answer with \system.
The most common problem was that additional data was needed that is currently not included in \system. For example, several participants wanted to see newer data than we provided (P2 \textit{``I would need to have more data after 2011.''}) or additional data we had not extracted (yet) from the blockchain (P2: \textit{``I would need to know balances\dots and more info about transactions. Are those P2PKH or P2SH\dots''}). 

In addition to their effectiveness answering their research questions, we also wanted to know if participants identified data that was new and surprising to them. Participant~P1 \emph{``found specific outliers and \ldots was amazed by the amounts of BTC exchanged.''} This participant~(P1) also mentioned that \emph{``there [were] a lot of one time entities''} and wondered if \emph{``they [were] one time users or errors on entity clustering.''}
Participant~P6 replied that \emph{``[c]oncerning Bitminter} [a supposed pool of miners]\emph{, there is always 1 outgoing address. I expected that block rewards were spread across all miners of the pool.''} Participant~P3 found that especially the clustering offered new and surprising results. Despite the inability of \system to help with several of the participants' questions, it is interesting to note the positive responses to the user experience-related questions reported above. It is likely the ability to see data in new ways and perhaps discover new aspects of the Bitcoin blockchain that led to positive responses about the system.

\section{Discussion}\label{sec:discussion}



\system's preliminary evaluation revealed the usefulness of its approach for answering questions related to entities rather than addresses---as well as questions that require larger-scale instead of small-scale analyses only as other tools provide. Still, we identified two major limitations related to our approach. 
One refers to the correctness and expression power of the data, the other to our analysis approach in the current form.

\noindent\textbf{Data.} The first limitation when it comes to the data we used is the role of entity building. 
We cannot expect the entities to be entirely correct and it is hard to estimate the quality of this information because the amount and quality of ground truth data are not sufficient. 
Tagging with information about known entities, however, can be useful to make sense of the activity of specific entities. 
However, most tags refer to the big players only, as Lischke and Fabian~\cite{Lischke:2016} pointed out in their analysis of the first four years of Bitcoin: ``Even though 54\% of all transactions could be related to a business tag and category, only 1.5\% of them are not associated to the major businesses SatoshiDICE, Mt.Gox, or Deepbit.''
Looking at the compression ratio of using entities vs.\ using addresses, during 2009--2001, the ratio is only $2/3$.  However this ratio is misleading because most of the entities (85\%) were ``one-timers'', something that an address-based analysis would not have revealed so easily.

\noindent\textbf{Approach.} The goal of \system's development was to support exploratory analysis of entity activities. Technically, this approach requires flexible computation of activity measures, which causes limitations with respect to scalability. While we tried to keep data processing times in the range of a few seconds the main bottleneck is the on-the-fly clustering that can take minutes in the worst case (if many entities are clustered at once). Here, a progressive clustering strategy may help where entities are not clustered all at once but intermediate clustering results are being displayed~\cite{fekete2016progressive}.

A major challenge in our approach is the aggregation of entities, which is computationally expensive and has a large memory footprint in our implementation. In the future, data providers could perform the aggregation operation. Google, for example, has recently started to distribute blockchain data maintained directly from its cloud infrastructure~\cite{GoogleBigQuery}. When aggregated entities become recognized as important for analysis, cloud data providers might similarly make regularly updated clustered data easily available.

Participants asked for additional data  and certainly further information could be added to enrich analyses. 
For instance, other approaches (\autoref{sec:relatedwork}) included information about geographic locations of addresses or entities. 
We did not include this information in \system\ because of its questionable quality. 
Geographic information is usually derived from the IP address of the last gateway before access of the blockchain. 
This is highly vague because the last address before the blockchain does not have to belong to a Bitcoin user. 
In addition, only a low fraction of transactions can be geotagged with existing strategies (e.g., in an analysis by Lischke and Fabian~\cite{Lischke:2016} around 1.6\% of all transactions). 
There are other strategies such as recursive scanning of Bitcoin nodes~\cite{Bitnodes:2018} but in general, geographic information about Bitcoin nodes is not a reliable source for analyses.

\noindent\textbf{Implications.} Based on \system\ and future extensions, we are hopeful that more research can reveal how cryptocurrencies are used in the real-world, beyond the vague and sensational information reported in the news only based on aggregated statistics or anecdotes from police reports. Yet, \system\ relies heavily on the capability of aggregating addresses meaningfully, which can become harder due to the development of complex mechanisms to increase anonymity in Bitcoin. 
As a reaction to anonymity issues, there are \emph{mixing services} or \emph{tumblers} (e.g., Coinmixer~\cite{CoinMixer:2018}) that redistribute Bitcoins in a pool of users so that the sending addresses of a transaction are not directly linked to the target addresses anymore. 
Past research has shown that this strategy can be an effective way to increase anonymity~\cite{moser2013anonymity}.  We believe these mechanisms will merely increase the uncertainty related to certain entities, but because our goal is not to detect crime, \system\ might still be usable for exploring transactions not related to criminal activity.

\section{Conclusion and Future Work}
We presented \system, a tool for exploratory visual analysis of activity on the Bitcoin blockchain. It consists of a data back end and a graphical browser-based interface as a front end comprising five linked views (filter view, tree view, cluster view, entity browser, and transaction view, see \autoref{fig:teaser}).

Our main novel contribution is to make in-depth visual exploration of Bitcoin entity activity possible, lowering the threshold for analyses especially for analysts without the technical background to prepare and handle the data for analysis. In addition, instead of providing the raw data, we facilitate analyses based on aggregated addresses (\emph{entities}) over large-scale time periods. An important part of our workflow involves systematic and reproducible classification of entity activities using filtering coupled with a tree representation of to characterize groups of entities. Human-assisted clustering helps to group entities with similar activity. Starting with large scale analyses it is possible to drill down and retrieve detailed information on single entities and display their transactions on a timeline.

In two use cases we demonstrated how \system can help characterize entity activity in two different settings. The first use case focused on the early years of Bitcoin and compared the amount of entities with only a single transaction per year. Analysis of the most active entities showed that the concentration of activity already took place in the first years. The second use case dealt with a specific event, the halving day in 2016. A closer look at the months before and after this day revealed that about 85\% of the active miners before the halving day did not continue in the month after this event. This revealed the impact of the event on mining behavior. Both use cases demonstrate that \system makes rather complex analyses relatively straightforward for the analyst, compared to existing approaches.

During a workshop with Bitcoin experts we learned that several research questions they had could easily be answered using the \system tool (e.g., about trends and outliers in activities, or mining behavior). 
Questions regarding temporal trends could not be answered (e.g., seasonality in the use) and pointed out a limitation of the approach.  
Ratings concerning \system's usability (confidence, ease-of-use, learnability) were predominantly positive. 
In particular, the integration of functionality in the GUI obtained high ratings and overall, five out of six experts claimed they would like to use \system more frequently.

Several limitations of our approach stem from the limited expressiveness of the data. Aggregation of addresses to entities provides a new perspective but introduces uncertainty that cannot be quantified reliably at this stage due to missing ground truth information. One way to decrease the uncertainty would be to include more external information such as tagging of entities. However, at the moment tagging has limited expression power.

The workshop showed that the most important extension of our work is a more convenient comparison of temporal patterns. An additional view could be integrated into the workflow, for instance in form of a radial chart or similar to facilitate comparison of activity patterns over different time periods. Another useful extension would be to provide similarity search, i.e., suggestion of entities that are similar to a specific entity of interest. Lastly, future work will be to add the capability to track addresses and individual entities by integrating the functionality we demonstrated in a separate tool called the \emph{Blockchain Entity Explorer}~\cite{isenberg:hal-01658500}.



\bibliographystyle{IEEEtran}
\bibliography{bitconduite}

\begin{IEEEbiography}
[{\includegraphics[width=1in,height=1.25in,clip,keepaspectratio]{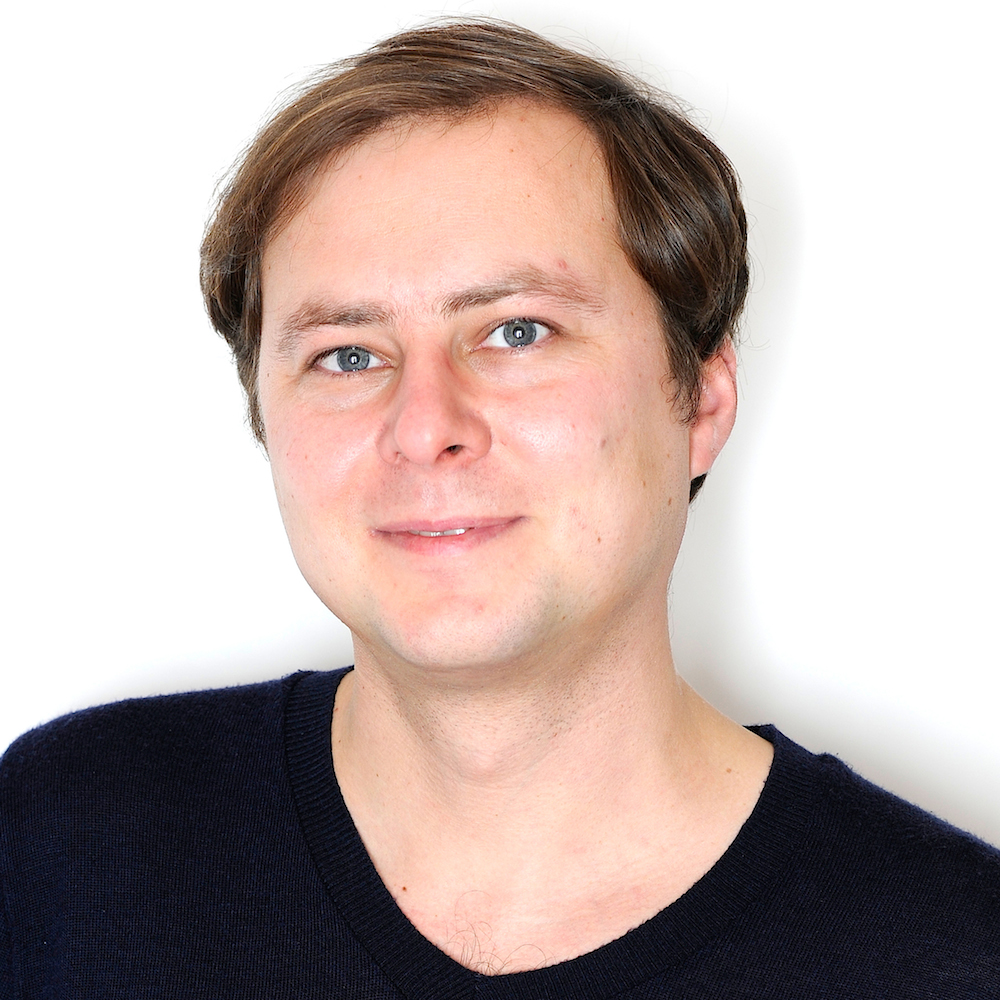}}]
{Christoph Kinkeldey}
is a Post Doctoral Researcher with the Aviz team at Inria. He received his Ph.D. in Geoinformatics in 2015 from HafenCity University Hamburg and specializes in visual analytics and geovisualization. He is particularly interested in how uncertainty visualization can support people to gain a better understanding of complex information.
\end{IEEEbiography}

\begin{IEEEbiography}
[{\includegraphics[width=1in,height=1.25in,clip,keepaspectratio]{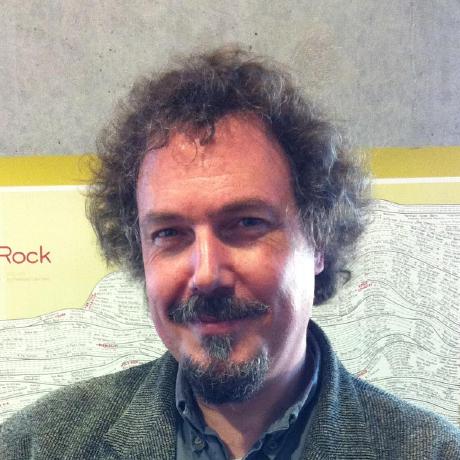}}]
{Jean-Daniel Fekete}
is the Scientific Leader of the Inria Project Team Aviz that he founded in 2007. He received his PhD in Computer Science in 1996 from University of Paris Sud, France, joined INRIA in 2002 and became Senior Research Scientist in 2006. His main research areas are Visual Analytics, Information Visualization and Human Computer Interaction. He is a Senior Member of IEEE.
\end{IEEEbiography}

\begin{IEEEbiography}
[{\includegraphics[width=1in,height=1.25in,clip,keepaspectratio]{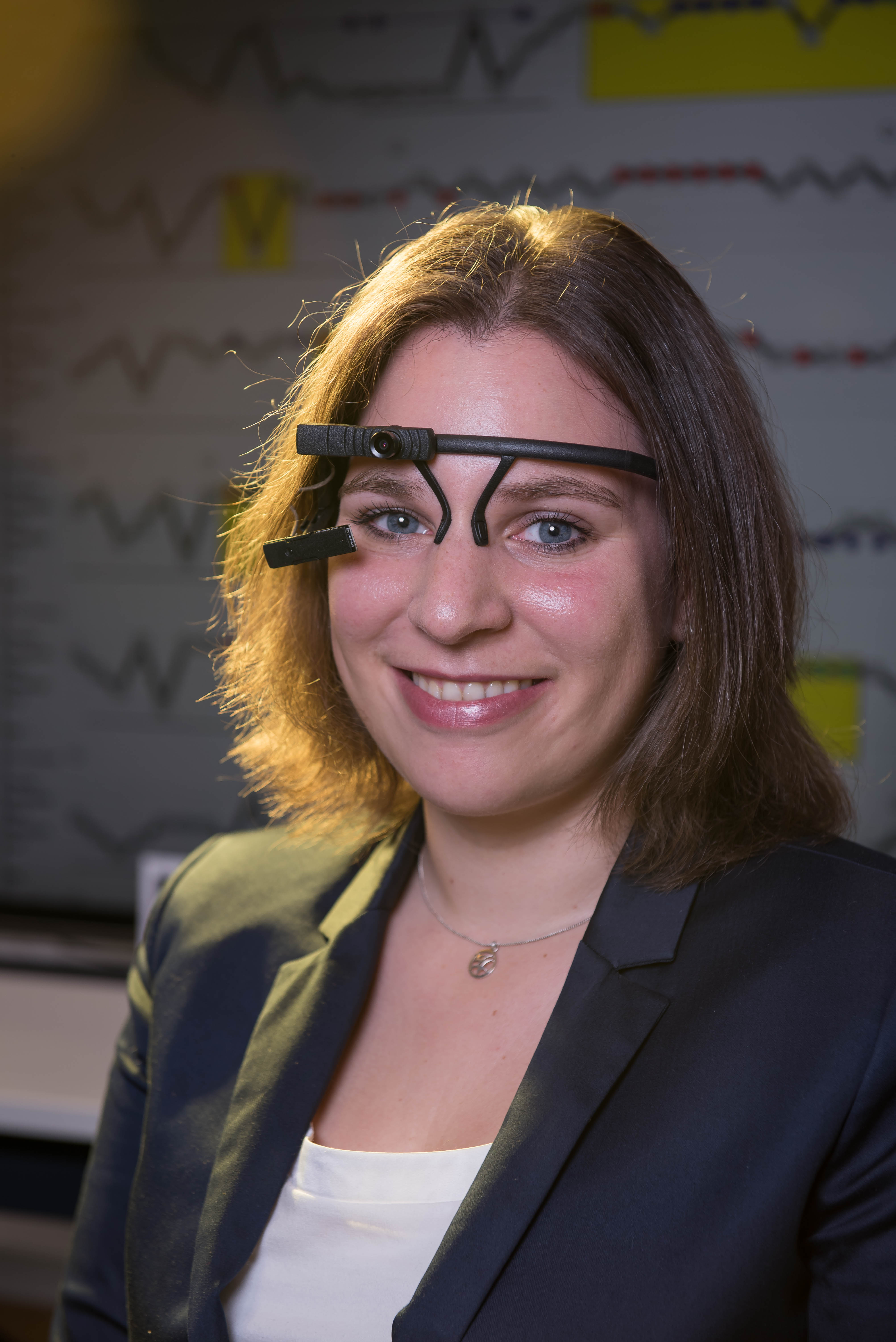}}]
{Tanja Blaschek}
is a Post Doctoral Researcher at Inria. Her main research areas are information visualization and visual analytics with a focus on evaluation, eye tracking, and interaction. She is interested in exploring how to effectively analyze eye tracking data with visualizations and the pervasive use of visualization on novel display technology like smartwatches. She received her Ph.D. in Computer Science from the University of Stuttgart, in 2017.
\end{IEEEbiography}

\begin{IEEEbiography}
[{\includegraphics[width=1in,height=1.25in,clip,keepaspectratio]{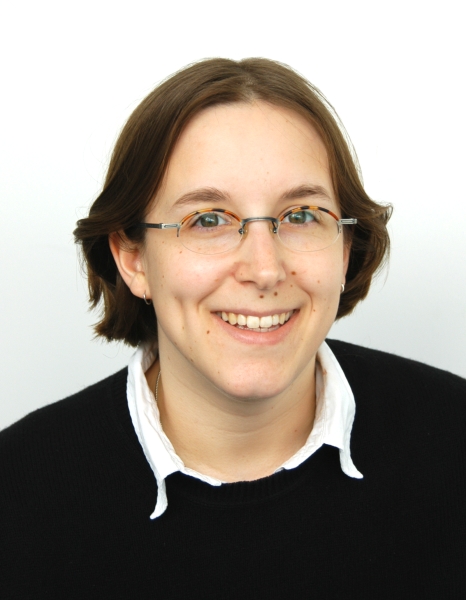}}]
{Petra Isenberg}
is a research scientist at Inria, France in the Aviz team. Her main research areas are information visualization and visual analytics with a focus on collaborative work scenarios, interaction, and evaluation. She is interested in exploring how people can most effectively work when analyzing large and complex data sets---often on novel display technology such as small touch-screens, wall displays, or tabletops. 
\end{IEEEbiography}

\end{document}